\documentclass{article}

     \PassOptionsToPackage{numbers, compress}{natbib}

 \usepackage[preprint]{neurips_2024}

\usepackage{algorithm}
\usepackage[noend]{algpseudocode}
\usepackage[utf8]{inputenc} 
\usepackage[T1]{fontenc}    
\usepackage{hyperref}       
\usepackage{url}            
\usepackage{booktabs}       
\usepackage{amsfonts}       
\usepackage{nicefrac}       
\usepackage{microtype}      
\usepackage[dvipsnames]{xcolor}         
\usepackage{graphicx}
\usepackage{doi}
\usepackage{subcaption}
\usepackage{listings}
\usepackage{array}
\usepackage{adjustbox}
\usepackage{multirow}
\usepackage{amsmath}
\usepackage{pifont}
\usepackage[flushleft]{threeparttable}
\usepackage{arydshln}
\usepackage{bm}
\usepackage{wrapfig}
\usepackage{booktabs}

\usepackage{diagbox}
\usepackage{placeins}

\usepackage{xr}
\makeatletter
\newcommand*{\addFileDependency}[1]{
\typeout{(#1)}
\@addtofilelist{#1}
\IfFileExists{#1}{}{\typeout{No file #1.}}
}\makeatother

\title{OpenQDC: Open Quantum Data Commons}
\date{June 2024}
\author{%
    Cristian Gabellini \thanks{Reach out at [cristian, prudencio]@valencelabs.com}\\
  Valence Labs\\
  \And 
  Nikhil Shenoy \\
  Valence Labs\\
  \And
  Stephan Thaler \\
  Valence Labs\\
  \And
  Semih Canturk \\
  Valence Labs, MILA\\
  \And 
  Daniel McNeela \\
  University of Wisconsin-Madison\\\\
  \And 
  Dominique Beaini \\
  Valence Labs, MILA, Montreal University \\
    \And 
  Michael Bronstein \\
  Oxford University \\
  \And
    Prudencio Tossou \footnotemark[1]\\
  Valence Labs \\
}

\begin{document}

\hypersetup{
pdftitle={OpenQDC},
pdfsubject={cs.LG},
pdfauthor={},
pdfkeywords={}
}
\maketitle

\begin{abstract}
Machine Learning Interatomic Potentials (MLIPs) are a
highly promising alternative to force-fields for molecular dynamics (MD) simulations, offering precise and rapid energy and force calculations. However, Quantum-Mechanical (QM) datasets, crucial for MLIPs, are fragmented across various repositories, hindering accessibility and model development. We introduce the openQDC package, consolidating 37 QM datasets from over 250 quantum methods and 400 million geometries into a single, accessible resource. These datasets are meticulously preprocessed, and standardized for MLIP training, covering a wide range of chemical elements and interactions relevant in organic chemistry. OpenQDC includes tools for normalization and integration, easily accessible via Python. Experiments with
well-known architectures like SchNet, TorchMD-Net, and DimeNet reveal challenges for those architectures and constitute a leaderboard to accelerate benchmarking and guide novel algorithms development. Continuously adding datasets to OpenQDC will democratize QM dataset access, foster more collaboration and innovation, enhance MLIP development, and support their adoption in the MD field.

\end{abstract}

\section{Introduction}
\label{sec:intro}
Molecular Dynamics (MD) simulations are invaluable in drug and material discovery, providing insights into the dynamic behavior and finite-temperature properties of bio-organic and inorganic molecules. They enhance our understanding of protein folding, allosteric sites, binding mechanisms, complex formation, kinetics \cite{romanowska2015computational, bernetti2017protein, bruce2018new, ribeiro2018kinetics}, protein-ligand binding free energy estimation \cite{kerrigan2013molecular}, and more. Their accuracy depends on the potential energy function, which defines intra- and intermolecular forces. \textit{Ab-initio} MD simulations use potentials from first-principles quantum mechanics (QM) methods such as Coupled Cluster or Density Functional Theory (DFT). However, \textit{ab-initio} calculations for large molecular systems are computationally intractable, even for the most powerful supercomputers. Therefore, MD simulations of bio-organic systems necessitate a trade-off between accuracy and speed, making empirical force fields, which prioritize computational speed over accuracy, the preferred method for studying biological systems \cite{karplus2005molecular, sinha2022applications}.

Two alternatives to empirical force fields are semi-empirical (SE) quantum mechanics \cite{gfn2, dftb3, pm6} and machine learning interatomic potentials (MLIPs) \cite{behler2007generalized, smith2017ani, khorshidi2016amp, unke2018reactive, tholke2021equivariant, kovacs2021linear, artrith2017efficient, takamoto2022towards}.
SE methods are lower-cost and feature moderate accuracy compared to DFT, but remain too expensive to perform long-timescale MD simulations necessary in biophysics.
MLIPs, trained using QM data (DFT or Coupled Cluster), expedite energy and force calculations compared to SE methods while maintaining precision comparable to QM. Recent advances in machine learning (ML) promise potentials with unprecedented speed, accuracy, scalability, and universality, and the field of MLIP modeling is rapidly growing, with numerous novel geometric deep learning architectures, physics-inspired architectures, and physical descriptors being developed \cite{gasteiger2020directional, gasteiger2021gemnet, liu2021spherical, satorras2021n, chen2022universal, schutt2021equivariant, thomas2018tensor, qiao2021unite}. Notable MLIP examples include ANI~\cite{ani1}, TorchANI~\cite{gao2020torchani}, AIMNet\cite{aimnet, aimnet2}, and MACE-OFF\cite{kovacs2023mace} potentials. 

Despite progress in MLIP modeling, several practical issues hinder their widespread development and adoption. First, large amounts of QM data are needed for training, which is expensive and limited in chemical coverage and size,
representing a major bottleneck for MLIP development.
Second, limited generalization to unseen chemical spaces arises from data limitations and architectural expressivity \cite{Fu2022, Thaler2022force, shen2018molecular}. Third, MLIPs are currently slower and less stable compared to empirical force fields.
These issues indicate that more research is needed for ML methods to fulfill their potential in QM data modeling. Given the chemistry and QM expertise required to understand QM datasets, lowering the barrier for ML experts and fostering the right research environment is essential.

In Natural Language Processing (NLP) and Computer Vision (CV), readily available and high-quality datasets such as ImageNet~\cite{imagenet_cvpr09}, CIFAR~\cite{cifar}, Cityscapes~\cite{cityscapes}, WikiText103~\cite{merity2016pointer} and CC-100~\cite{cc100-0, cc100-1}, have driven advancements. These datasets, packaged in libraries such as TorchvisionDataset, catalyze research, foster cross-disciplinary collaboration, and lower the learning curve for non-experts. In contrast, MLIP modeling lacks standardized, ready-to-use datasets for benchmarking algorithms and fostering innovation. Existing QM datasets span various methods and domains of the chemical space, but are dispersed across different repositories and formats.
This hinders adoption and utility in molecular simulations and quantum calculations, impedes cross-collaboration among physicists, chemists, ML scientists, and other fields, and limits the effectiveness and progress of MLIP research. Enhancing data accessibility is crucial for advancing MLIP development and adoption. Similar resources are needed for MLIP innovation.

In this paper, we address this issue by presenting Open Quantum Data Commons (OpenQDC), a comprehensive dataset repository and library of ML-ready datasets designed for MLIP algorithm development and training. Our contributions are summarized as follows:
\begin{itemize}
    \item We have gathered publicly available QM datasets, consolidating them into the OpenQDC repository for enhanced accessibility, MLIP training, and collaboration. It includes nearly 40 datasets, covering 400 million geometries across 70 atom types and 250+ QM methods. The repository contains both potential and interaction energy datasets. The former is essential for training MLIPs, and the latter for training scoring models for drug-target interactions (DTI), drug-drug interactions (DDI), and protein-protein interactions (PPI).
    \item We have open-sourced this repository through the OpenQDC Python library, where every dataset is ready to use with a single line of code. The package offers tools to properly normalize and convert energies and forces during MLIP training, and makes the addition of new datasets more convenient. This represents a turning point for MLIP modeling akin to the publication of HuggingFace datasets \cite{lhoest-etal-2021-datasets}, Torchvision Datasets \cite{torchvision2016}, or TDC \cite{Huang2021tdc}.
    \item We have benchmarked and reported the performance of several well-known architectures such as TorchMD-Net, DimeNet, and SchNet on single-task MLIP training. This effort will save significant amounts of time and computational resources for the development and benchmarking of novel architectures.
\end{itemize}

\section{Related works}
\label{sec:relatedworks}
\paragraph{QM data repositories:}
There have been many attempts to centralize QM datasets. Notable repositories are Quantum Machine \cite{QuantumMachine}, Quantum Chemistry Archive (QCArchive)\cite{smith2021molssi}, and ColabFit Exchange\cite{vita2023colabfit}. Quantum Chemistry hosts 17 datasets and benchmarks aimed at accelerating MLIP development. These datasets are available in different archive formats and require additional preprocessing for ML training. QCArchive offers a community-wide repository with curated datasets in HDF5 and CSV formats, covering interaction energy, thermochemical, and conformational data of approximately 39 million bio-organic molecule conformers across 24 datasets. This facilitates QM data generation and sharing for large-scale MLIP construction, QM property prediction, and methodology assessment. ColabFit Exchange includes 372 datasets, mostly in XYZ format, encompassing 180 million conformers of inorganic molecules, with a focus on materials science. It is also worth mentioning Begdb~\cite{begdb} and Cuby~\cite{Cuby} which are smaller repositories of QM interaction datasets which cover different formats.

Both QCArchive and ColabFit aim to enhance the Findability, Accessibility, Interoperability, and Reusability (FAIR) of QM data. However, they remain relatively unknown among non-experts in QM and force field modeling and rely primarily on web interfaces rather than programmatic interfaces optimized for ML purposes. In contrast, OpenQDC offers both a website (to be release upon acceptance) and a programmatic interface to the repository, along with tools to enhance FAIR principles beyond previous capabilities.

\paragraph{ML-ready QM datasets:} 
A few QM datasets are ML-ready and accessible in ML libraries such as TDC\cite{TDC}, MoleculeNET\cite{wu2018moleculenet}, Torch Geometric Datasets\cite{pyg}, Deep Graph Library\cite{dgl}, and TorchDrug\cite{zhu2022torchdrug}. These are primarily used to benchmark novel architectures for 2D graphs or 3D atomistic systems, and occasionally to train MLIPs. Notable datasets include \textsc{QM7}\cite{qm7}, a subset of GDB-13 with 7,165 molecules having at most seven CNOS atoms and their atomization energy, and other QM properties; QM7b extends QM7 with 13 additional properties for 7,211 molecules; QM8\cite{qm8} provides electronic properties for 21,786 molecules with up to eight CNOF atoms; QM9\cite{qm9} includes QM properties from a single DFT method for 133,885 organic molecules with up to nine non-hydrogen atoms; and MD17\cite{gdml} offers \textit{ab-initio} molecular dynamics trajectories for training MLIPs. These datasets cover limited chemical space and small molecules. A larger dataset, \textsc{PCQM4M(v2)}\cite{pubchemqc_pm6}, includes the HOMO-LUMO gap and equilibrium structure energies for approximately 3.6 million molecules and is accessible through the Open Graph Benchmark library (OGB)\cite{hu2021ogb-lsc}, challenging the community to infer 3D information from 2D graphs during inference.

\paragraph{Libraries for MLIP research and development:} 
Many libraries have emerged to accelerate MLIP development and MD usage. DeePMD-kit~\cite{wang2018deepmd} integrates with TensorFlow and classical/path-integral MD packages, implementing Deep Potential models with MPI and GPU support. DP-GEN~\cite{zhang2020dp} aids in generating reliable deep learning models by facilitating QM dataset generation, training, and active learning. PiNN~\cite{shao2020pinn} and KLIFF~\cite{wen2022kliff} offer tools for building neural network-based MLIPs and using them in MD. Finally, MD simulation codes written in ML frameworks such as JaxMD~\cite{schoenholz2020jax} and TorchMD~\cite{doerr2021torchmd} facilitate research at the interface between MLIP development and MD applications.

More recent and state-of-the-art architectures are often available in specialized libraries. TorchANI~\cite{gao2020torchani} is a PyTorch library for training/inference of the ANI model series. TorchMD-NET 2.0 provides efficient implementations of several MLIP architectures (TorchMD-NET~\cite{pelaez2024torchmdnet}, TensorNet~\cite{simeon2023tensornet}, ET~\cite{tholke2021equivariant}, CG-NNP~\cite{Majewski2023}). The Neural Force Field library includes architectures like SchNet~\cite{schutt2017quantum, schutt2017schnet, schutt2018schnet, schutt2018schnetpack}, DimeNet~\cite{gasteiger2020directional}, PaiNN~\cite{schutt2021equivariant, axelrod2023thermal}, and DANN~\cite{axelrod2022excited}. The MACE library~\cite{batatia2022mace, batatia2022design} provides high-order equivariant message passing architectures and pretrained models such as MACE-OFF~\cite{kovacs2023mace} and MACE-MP~\cite{batatia2023foundation}. Similarly, the Allegro and NeuqIP libraries offer implementations and pretrained models based on the e3nn framework \cite{e3nn_paper} for building E(3)-equivariant NNs.

Despite these advancements, there is still a gap in data accessibility that impedes the progress of MLIP research. OpenQDC aims to fill this gap by providing standardized, ready-to-use datasets, facilitating research on more performant MLIP architectures by the machine learning community.

\section{OpenQDC Datasets}
\label{sec:datasets}
OpenQDC consists of a collection of previously published and publicly available QM datasets. Table~\ref{tab:data_stats} shows an overview of the statistics of those datasets and proves that OpenQDC is the largest collection of QM data in term of number of geometries, energy and force labels, and atom statistics.  These datasets mainly differ by their chemical coverage, the types of properties computed and their 
QM levels of theory A/B, which are the specific combinations of QM method A and basis set B used to compute the electronic structure and properties of molecules.
Different levels of theory vary in their accuracy and computational cost, influencing the reliability of the predicted results making our collection a multi-fidelity one that can support ML research in multi-fidelity approaches. The datasets are divided into two categories: potential energy datasets and interaction energy datasets. The former includes datasets where QM labels represent the potential energies and optionally the derived atomic forces of 3D atomistic systems. The latter includes datasets where the labels represent the interaction energies computed between two or more 3D atomistic systems. Programmatic access to these collections is provided through the OpenQDC Python package, which includes a comprehensive set of tools to support data processing and MLIP training.
Below, we describe each collection. 

\subsection{Potential Energy Collection}
This collection includes datasets primarily used for MLIP training, providing the potential energy of 3D systems and optionally the forces on each atom. These datasets are well-known in the QM literature but less familiar to ML experts developing new architectures for advancing the MLIP field. Although not exhaustive, this collection includes recent or sufficiently large datasets for MLIP training, focusing on bio-organic systems relevant to drug discovery, an area still underexplored compared to material discovery. Visualizations for this collection are available in the Appendix~\ref{appendix:potential_visuals}

\begin{table}[h]\centering
\caption{Statistics of the datasets in both collections:  the number of conformers(\# conf.), energy labels (\# E), force labels (\# F), atom types (\# Atom type) and the minimum-manximum atoms (Atom Min/Max) of geometries  in every dataset. 
}\label{tab:data_stats}
\scriptsize
\adjustbox{width=\textwidth}{
\begin{tabular}{p{0.08\linewidth}lllp{0.04\linewidth}p{0.06\linewidth}|p{0.1\linewidth}lllp{0.04\linewidth}p{0.06\linewidth}} 
\toprule
\bf{Dataset}  & \bf{\# conf.} & \bf{\# E } & \bf{\# F }  & \bf{\# Atom type} & \bf{ Atom Min/Max} & \bf{Dataset} & \bf{\# conf.} & \bf{\# E } & \bf{\# F }  & \bf{\# Atom type} & \bf{ Atom Min/Max}   \\
\midrule
ANI-1  & 22,057,374 &1 &0 & 4 & 2/26& ANI-1x & 4,956,005 &8 &2 & 4 & 2/63\\
ANI-1ccx & 489,571 &4 &0 & 4 & 2/63 & ANI-2x & 9,651,712 &1 &1 & 7 & 2/63 \\
COMP6 &101,352 &7 &1 & 4& 6/312 & GDML &3,875,468 &3 &3 & 4& 9/24\\
GEOM &33,078,483 &1 &0 &13 & 3/181 & ISO17 &640,982 &1 &1 & 3& 19/19 \\
MD22 &223,422 &1 &1 & 4 & 42/370 & Molecule3D & 3,899,647 &1 &0 & 36 & 1/137 \\
MultixcQM9 &133,631 &229 &0 & 5 & 3/29 & NablaDFT  & 1,275,340 &1 &1 & 8 & 8/57 \\
OrbNet D. &2,338,889 &2 &0 & 17 & 2/74 & Pub. PM6  & 189,890,155 &1 &0 & 70 & 1/215 \\
Pub. B3lyp & 85,915,773 &1 &0 & 70 & 1/215 & QM7 &7,165 &1 &0 & 5 & 3/7 \\
QM7-X &4,195,192 &2 &1 &6 & 4/23 & QM8 & 21,786 &2 & 0 & 5 & 3/8 \\
QM9 &133,885 & 1 & 0 & 5 & 3/9 & Qmugs & 1,992,984 &2 &0 & 10 & 4/228\\
RevMD17 & 999,988 &1 &1 & 4 & 9/24 & SN2 React. & 452,709 &1 &0 & 6 & 2/6\\
Sol. Prot. & 2,731,180 &1 &1 & 5 & 2/120 & Spice & 1,110,165 &1 &1 & 15 & 2/110 \\
SpiceV2 & 2,008,628 &1 &1 & 17 & 2/110 & tmQM& 86,665 &1 &0 & 44 & 5/569 \\
Transition1x & 9,654,813 &1 &1 & 4 & 4/23 & WaterClusters & 4,464,740 &1 &0 & 2 & 9/90\\
\midrule
\textbf{Potential Total}& 386,387,704&278&16&70 &1/370 &&&& \\
\midrule
DES370K & 370,959 &14 &0 & 20 & 2/44 & DES5M & 4,955,938 &17 &0 & 14 & 2/34 \\
DESS66 & 66 &17 &0 & 4 & 6/34 & DESS66x8 & 528 &17 &0 & 4 & 6/34 \\
Metcalf & 13,415 &5 &0 & 6 & 12/41 & X40 & 40 &5 &0 & 9 &7/25 \\
L7 & 7 & 8 &0 & 4 & 48/112 & Splinter & 1,677,830 &20 &0 & 10 & 2/51 \\
\midrule
\textbf{Interaction Total}& 7,018,783&-&0&20 &2/112 &&&& \\
\bottomrule
\end{tabular}
}
\end{table}

\textbf{ANI-1~\cite{ani1}} comprises molecular energies at the $\omega$B97X39/6-31G(d) level of theory for $17.2$ million conformations generated from $58k$ small molecules 
from GDB-11\cite{GDB}. The geometries were sampled by calculating the normal modes and then randomly perturbing the equilibrium structures. 

\textbf{ANI-1x~\cite{comp6_ani1x, ani1x_ani1ccx}} was constructed using active learning and contains DFT data for $5$ million conformations of molecules averaging $15$ atoms. It adds new QM calculations to improve the accuracy of MLIPs trained with ANI data and despite its smaller size, ANI-1x-trained MLIPs outperform those trained on ANI-1, making it more suitable for training transferable and extensible MLIPs.

\textbf{ANI-1ccx~\cite{ani1ccx, ani1x_ani1ccx}} includes $500k$ conformers subsampled from the ANI-1x dataset and labeled using a high-accuracy CCSD(T)*/CBS method, enabling the training of MLIPs that can surpass DFT methods in accuracy.

\textbf{ANI-2x} was constructed using active learning from modified versions of GDB-11\cite{GDB}, CheMBL\cite{chembl}, and s66x8\cite{s66x8}. It adds three new elements (F, Cl, S) resulting in $4.6$ million conformers from $13k$ chemical isomers, optimized using the LBFGS algorithm and labeled with $\omega$B97X/6-31G*.

\textbf{COMP6~\cite{comp6_ani1x}} comprises six subsets (S66x8, ANI-MD, GDB7to9, GDB10to13, Tripeptides, and DrugBank), totaling $101k$ geometries from non-equilibrium molecular conformations with energies and forces calculated using the DFT $\omega$B97x59/6-31G(d) level of theory. These subsets cover diverse organic molecules and noncovalent interactions.

\textbf{GDML~\cite{gdml}} consists of $3.2$ million geometries from eight molecules sampled from \textit{ab-initio} MD trajectories at 500 K with a resolution of 0.5 fs. Energy and force labels were computed using the PBE+vdW-TS method. 

\textbf{GEOM~\cite{geom}} offers $37$ million molecular conformations for over $450k$ molecules, primarily generated using the semi-empirical extended tight-binding method (GFN2-xTB\cite{gfn2-xtb}) and labeled with the same method. A subset of $1,511$ molecules was labeled with high-quality DFT in an implicit water solvent.

\textbf{ISO17~\cite{iso17_schnet}} consists of $129$ molecules, each with $5,000$ conformational geometries from \textit{ab-initio} MD. Energy and force labels were computed using the PBE+vdW-TS method.

\textbf{MD22~\cite{md22}} includes four classes of biomolecules and supramolecules, ranging from a small peptide with $42$ atoms to a double-walled nanotube with $370$ atoms. Geometries were extracted from MD trajectories at $400-500$ K with a resolution of $1$ fs, with potential energies and atomic forces calculated at the PBE+MBD level of theory.

\textbf{Molecule3D~\cite{molecule3d}} contains precise ground-state geometries of approximately $4$ million molecules from the PubChem database, calculated using DFT at the B3LYP/6-31G* level of theory.

\textbf{MultiXCQM9~\cite{multixc_qm9}} re-labels QM9~\cite{ramakrishnan2014quantum} molecules with $76$ different DFT functionals and three basis sets per functional, totaling $228$ DFT methods. It also includes GFN2-xTB\cite{gfn2-xtb} energy calculations, enabling various MLIP tasks like transfer learning, delta learning, and multitask learning.

\textbf{NablaDFT~\cite{nabladft}} provides $12.6$ million conformations from $1.9$ million molecules from the MOSES dataset, covering a wide range of drug-like compounds. Energies, forces, and Hamiltonians were calculated using the Kohn-Sham method at the $\omega$B97X-D/def2-SVP level of theory.

\textbf{OrbNet Denali~\cite{orbnet}} consists of $2.3$ million conformers that cover a wide range of organic molecules, protonation and tautomeric states, non-covalent interactions, common salts, and counterions. Structures were optimized with GFN1-xTB, and DFT calculations with $\omega$B97X-D3/def2-TZVP were conducted. This dataset enables reliable MLIP training across various chemical environments.


\textbf{PubChemQC PM6~\cite{pubchemqc_pm6}} consists of 221 million geometries optimized and labeled with SE PM6\cite{pm6}. They were  obtained from 91.6 million molecules cataloged in PubChem\cite{pubchem} and include neutral states, cationic, anionic, and spin-flipped electronic states for approximately half of the molecules. It is the largest collection of QM calculations on drug-like molecules, covering an extensive chemical space, and is expected to advance MLIP development for broad generalization.


\textbf{PubChemQC B3LYP/6-31G*//PM6~\cite{pubchemqc_b3lyp}} includes 86 million energy calculations using both the DFT level of theory B3LYP/6-31G$^*$ and the semi-empirical PM6 method\cite{pm6}. Similar to PubChemQC PM6, the geometries were obtained from molecules cataloged in PubChem\cite{pubchem} and cover the same chemical space. All geometries were optimized by the PM6 method before single-point calculations, and the use of DFT is expected to enhance the accuracy of MLIPs.

\textbf{QM7~\cite{qm7}, QM8~\cite{qm8}, QM9~\cite{qm9}} are part of a series aimed at providing comprehensive QM data on the building blocks that compose most organic molecules. They respectively cover molecules having 7, 8, and 9 heavy atoms and use various levels of theory. 

\textbf{QM7-X~\cite{qm7x}} includes $4.2$ million equilibrium and non-equilibrium structures of small organic molecules with up to seven heavy atoms. It covers constitutional/structural isomers, stereoisomers, and their conformational variations. Structure generation was performed using DFTB3+MBD, and QM calculations were done at the tightly converged PBE0+MBD level of theory. QM7-X provides a systematic and extensive dataset, enabling greater MLIP robustness due to its exhaustive sampling. 

\textbf{QMugs~\cite{qmugs}} is a collection of $2$ million conformers of $665K$ bioactive and pharmacologically relevant molecules extracted from ChEMBL\cite{chembl}. All geometries were generated with Rdkit and optimized using the semi-empirical GFN2-xTB method\cite{gfn2-xtb}. Energy and force calculations are provided using both GFN2-xTB and DFT ($\omega$B97X-D/def2-SVP) levels of theory. QMugs features significantly larger molecules than most other collections, and facilitates the development of MLIPs that can learn from different levels of theory.

\textbf{RevMD17~\cite{revMD17}} is a revision of MD17\cite{md17}, which contains numerical noise and autocorrelated samples, making it unsuitable for MLIP training. RevMD17 addresses the noise issue
to facilitate the benchmarking of force and energy predictions in MD simulations.

\textbf{SN2 Reactions~\cite{sn2_and_solvated}} probes chemical reactions of the type $X^- + H_3C-Y \rightarrow X-CH_3 + Y^- $ ($X, Y \in \{F, Cl, Br, I\}$), which are prototypical in biological systems. It consists of $452K$ geometries for high-energy transition regions, ion-dipole bound state complexes, and long-range ($>10$ Å) interactions between $CH_3X$ and $Y^-$ ions. Energies and forces are calculated at the DSD-BLYP-D3(BJ)/def2-TZVP level of theory. This dataset is valuable for building and benchmarking MLIPs that work far from equilibrium configurations and reactive systems.

\textbf{Solvated Protein Fragments~\cite{sn2_and_solvated}} contains all amons (hydrogen-saturated covalently bonded fragments) up to eight heavy atoms derived from proteins and their solvated variants up to 21 heavy atoms. It also includes randomly sampled dimer interactions of protein fragments and water clusters up to 40 molecules, with energies and forces for $2.7$ million geometries sampled with MD at 1000 K and optimized using the semi-empirical PM7 method. QM calculations were performed at the revPBE-D3(BJ)/def2-TZVP level of theory. This dataset is useful for learning MLIPs that generalize to many-body intermolecular interactions between protein fragments and water molecules.

\textbf{Spice~\cite{spice}} contains over 1.1 million conformations for a diverse set of small molecules, dimers, dipeptides, and solvated amino acids. It includes 15 elements, charged and uncharged molecules, and a wide range of covalent and non-covalent molecular systems. SpiceV2 \cite{spice_v2} extends to bulk water systems, solvated molecules, and two more chemical elements. Energies and forces are calculated at the $\omega$B97M-D3(BJ)/def2-TZVPPD level of theory, making Spice a valuable resource for training MLIPs relevant to simulating drug-like small molecules interacting with proteins.

\textbf{tmQM~\cite{tmqm}} has $86K$ mononuclear complexes of various organic ligands and 30 transition metals (3d, 4d, and 5d from groups 3 to 12). They were extracted from the Cambridge Structural Database\cite{groom2016cambridge}, were optimized at the GFN2-xTB\cite{gfn2-xtb} level. Quantum properties were computed at the DFT(TPSSh-D3BJ/def2-SVP) level. Given its composition, tmQM is useful for understanding metal complexes and building MLIPs with applications in catalysis, organic synthesis, and metalloprotein binding.

\textbf{Transition1x~\cite{transition1x}} contains 9.6 million DFT ($\omega$B97x/6-31$\cdot$G(d)\cite{chai2008systematic, ditchfield1971self}) forces and energy calculations of geometris along $~10k$ organic reaction pathways. These reactions involve up to 7 heavy atoms (C, N, O) and up to six bond changes, where bonds are breaking and forming between all combinations of heavy atoms. Transition1x was generated by running Nudged Elastic Band (NEB)\cite{sheppard2008optimization} calculations, followed by DFT calculations on intermediate states. This dataset is essential for developing MLIPs that operate far from equilibrium configurations and reactive systems.

\textbf{WaterClusters~\cite{watercluster_3_25}} consists of $\sim 3.2$ million local minima within 5 kcal/mol of the putative minima for water clusters of varying sizes ($n \in [3, 25]$). Samples were generated using a global optimization procedure with an ab initio-based flexible, polarizable Thole-Type Model (TTM2.1-F\cite{fanourgakis2006flexible}) interaction potential for water. This dataset spans diverse hydrogen-bonding networks, enabling MLIP development for water scaling from small clusters to larger ones.

\subsection{Interaction Energy Collection}

This collection of datasets is crucial for modeling various types of molecular interactions, including drug-drug interactions (DDIs), drug-target interactions (DTIs), and protein-protein interactions (PPIs), and quantifying their strengths. Each dataset provides interaction energies and geometries for a range of molecular complexes, offering valuable data for training ML models aimed at understanding these interaction types.
Visualizations for this collection are available in the Appendix~\ref{appendix:interaction_visuals}

\textbf{DES370K~\cite{all_dess}} features $370K$ unique geometries for 3,691 distinct dimers composed of 392 neutral molecules and ions, including water and protein functional groups. These structures were obtained through radial profile scanning from QM-optimized conformers and MD simulations. The interaction energies were computed at the gold-standard CCSD(T)/CBS level of theory, incorporating single, double, and perturbative triple excitations. This dataset is primarily intended for parameterizing new exchange-correlation ML functionals and MLIPs.

\textbf{DES5M~\cite{all_dess}} contains 5 million additional unique geometries derived from the same types of dimers in DES370K. The geometries were generated from radial profiles starting from QM-optimized conformers or extracted from MD simulations. Interaction energies were obtained using the SNS-MP2 predictive model, a machine learning approach that provides results with accuracy comparable to coupled-cluster training data. This dataset enhances the chemical diversity available for training interaction MLIPs and developing new exchange-correlation ML functionals.

\textbf{DESS66 and DESS66x8~\cite{all_dess}}. The S66~\cite{s66} dataset consists of equilibrium geometries for 66 complexes formed from 14 monomers representing motifs and functional groups common in biomolecules. S66x8~\cite{s66x8} includes eight geometries along the dissociation curve for each of the 66 complexes in S66. Interaction energies for these datasets were measured using CCSD(T)/CBS, making them valuable for benchmarking interaction energy models.

\textbf{Metcalf~\cite{metcalf}} features hydrogen-bonded dimers involving N-methylacetamide (NMA) paired with 126 different molecules (46 donors and 80 acceptors). Optimized geometries for each monomer were obtained and paired with NMA in various spatial configurations to generate thousands of complexes. Interaction energies were computed using the SAPT0 functional. This dataset is particularly useful for developing ML functionals that understand hydrogen-bond interactions.

\textbf{Splinter~\cite{splinter}} consists of 1.7 million different configurations of over $9K$ unique molecular dimers involving $332$ distinct small molecular structures commonly found in proteins and their ligands. Most geometries were randomly sampled to cover each dimer’s potential energy surface, with some minimized to obtain local and global minima. This ensures accurate determination of the shape and depth of the energy wells while covering unfavorable and non-interacting regions. Interaction energies were computed using SAPT0 with two basis sets. This dataset is intended for training and testing various methods for calculating intermolecular interaction energies of DDIs and DTIs.

\section{The OpenQDC Library}
\label{sec:OpenQDC}
The OpenQDC library (\url{https://github.com/OpenDrugDiscovery/openQDC}) is designed to facilitate access and manipulation of QM datasets for training MLIPs. To build this library, we fetched the datasets from their respective sources and preprocessed them into a unified format to ensure fast loading, indexing, and batching. A significant challenge in this process was retrieving essential metadata (e.g., energy, distance, force units, and isolated atom energies) necessary for accurate data processing. For example, without proper distance units, radius-based architectures cannot be effectively trained. Similarly, isolated atom energies allow computation of the molecule's atomization energy, serving as a physics-motivated energy normalization method that respects the extensivity of potential energy. Most datasets lacked this information, so we computed these values for all QM methods in the repository. An API was then developed to make all datasets accessible via a single command line interface, featuring numerous functions for manipulation. The following sections discuss the shared data storage, loading, and features of the library.\\
\textbf{Dataset Storage:} For a dataset with $N$ geometries, $M$ atoms across all geometries, $n_e$ energy labels, and $n_f$ force labels, we use memory-mapped arrays of various sizes: ($M, 5$) for atomic numbers (1), charges (1), and positions (3) of individual geometries; ($N, 2$) for the beginning and end indices of each geometry in the previous array; ($N, n_e$) for the energy labels of each geometry, extendable to store other geometry-level QM properties such as HOMO-LUMO gap; ($M, n_f, 3$) for the force labels of each geometry, extendable to store other atom-level QM properties.

The memory-mapped files efficiently access data stored on disk or in the cloud without reading them into memory, enabling training on machines with smaller RAM than the dataset size and accommodating concurrent reads in multi-GPU training. This allows for very efficient indexing, batching and iteration. 

\begin{figure}[h]
    \centering 
    \includegraphics[width=\textwidth]{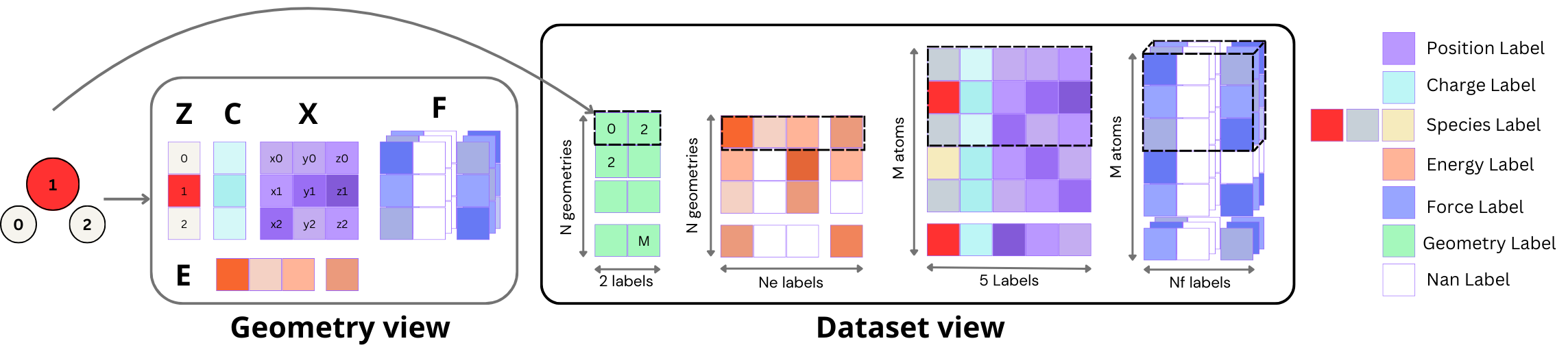}
    \caption{Overview of a dataset and its structure.}
    \label{fig:iid_qmugs_qm7x}
\end{figure}

\textbf{Dataset Loading:} Every dataset is accessible with a single line of code. For example, to load the Qmugs dataset:
\begin{lstlisting}
    from openqdc.datasets import QMugs
    dataset = QMugs(energy_unit="eV", distance_unit="ang", array_format="torch")  
\end{lstlisting}
The data class permits specification of the energy and atomic distance units (and consequently, the force unit) due to the storage of original units and behind-the-scenes conversions. This feature is critical for comparing quantities across many datasets. The original units and other crucial metadata, such as the QM methods used for computations, are stored in the class and can be accessed as properties. Additionally, users can choose the format for arrays returned from indexing and batching operations (e.g., torch, numpy, or jax) and decide whether to use isolated atom energies or atom type regression values to normalize the energy labels during the dataset object initialization.

\textbf{QM Method:} During data collection, we faced challenges with inconsistencies in QM method notation. The same QM method was often written differently, potentially confusing non-experts and causing issues when merging datasets. To solve this, we curated the QM notations and categorized them into Python Enums consisting of a functional, a basis set, and a correction method enabling the standardization of methods across datasets. This enables easier study of the effects of basis sets or correction terms for different functionals. OpenQDC covers 106 functionals, 20 basis sets, and 11 correction methods. Additionally, we computed the isolated atom energies for each QM method as properties of these structures and use them for physical normalization.

\textbf{Energy Normalization:} Special care needs to be taken in normalizing energy values given that the standard Z-transformation of the potential energy of molecules violates the extensivity property of the energy (i.e., that energy scales with the system size).  Conserving size extensivity is essential such that the MLIP can be applied to different systems than the training ones \cite{musaelian2023learning}. OpenQDC simplifies normalization by pre-computing atom type-specific shift and scale values for each dataset. Transforming potential energy to atomization energy by subtracting isolated atom energies is a physically interpretable and extensivity-conserving normalization method. Alternatively, we pre-compute the average contribution of each atom species to potential energy via linear or ridge regression, centering the distribution at 0 and providing uncertainty estimation for the computed values. Predicted atomic energies can also be scaled to approximate a standard normal distribution.

\paragraph{Functionalities:} All datasets in OpenQDC are iterable objects yielding molecular conformations and their labels, with additional functionalities:
\begin{itemize}
    \item \textbf{Label Statistics:} Tools for collecting statistics on energy and force labels, allowing further normalization within the MLIP by scaling predicted atomic energies.
    \item \textbf{Geometry Representation:} Physical descriptors (SOAP, ACSF, MBTR) for visualizing and comparing geometries within and between datasets.
    \item \textbf{Visualization:} Helpers for visualizing datasets.
    \item \textbf{Format Conversion:} Convert geometries or full datasets to ASE atoms objects or into XYZ and extXYZ files.
    \item \textbf{Splitting:} Tools for dataset splitting based on user-defined criteria.
    \item \textbf{Extensibility:} openQDC datasets can be created from common file formats.
\end{itemize}

\section{Experiments}
\label{sec:experiments}
\subsection{Setup}
\label{sec:setup}
Our initial benchmarking efforts focused on energy-matching to learn MLIPs of as many energy functionals as possible within our computational budget. We use three popular and fast to train architectures as starting points rather than state of the art architectures such as MACE~\cite{kovacs2023mace} or Allegro~\cite{musaelian2023learning}, which require far more computational resources than we had at our disposal. More precisely, we used TorchMDNet,  DimeNet and SchNet. TorchMDNet \cite{tholke2022torchmd}, is an equivariant Transformer (ET) architecture leveraging the use of rotationally equivariant features. DimeNet~\cite{gasteiger2020directional}, integrates directional message passing and spherical Fourier-Bessel representations for predictions invariant to permutation, translation, rotation and inversion. SchNet~\cite{schutt2017schnet}, is a continuous-filter convolutional network that respects rotation invariant for energy predictions and rotation-equivariant for energy-conserving force predictions.

A detailed list of hyperparameters for all architectures are provided in Appendix \ref{appendix:hyperparameters}. To train these models, we compute normalized energy labels $\Tilde{E}$ for each dataset and QM level of theory $j$ by subtracting the mean energy value of each atom type as computed by a linear model $\mu_{j,k}$, analogous to \cite{ramakrishnan2015big}. In addition, we scale the energy by the mean residual energy per atom $\sigma_j$:
\begin{equation}
    \Tilde{E}^\mathrm{QM}_{i,j} = \frac{1}{\sigma_j} \left(E^\mathrm{QM}_{i,j} - \sum_{k=1}^{K_i} \mu_{j,k}\right) \ .
\end{equation}
For all datasets, we performed a molecule/system-based split (80\%-10\%-10\% training-validation-test) evaluating the realistic case where the test molecule is not part of the training set.

\begin{table}[h]
\caption{ Results for the baseline archictectures. We report the MAE on the test set for both potential (\textit{top}) and interaction (\textit{bottom}) datasets and will be repeating the experiments over many seeds subsequently.}
\label{tab:results}
\adjustbox{width=\textwidth}{\begin{tabular}{lccc|lccc}
\toprule
Dataset  & TorchMDNet & DimeNet & SchNet & Dataset  & TorchMDNet & DimeNet & SchNet \\
\midrule
GEOM-Drugs&\textbf{0.267}&0.513&0.637&Molecule3D&\textbf{0.264}&0.286&0.441\\
GEOM-QM9&\textbf{0.078}&0.085&0.147&ISO17&0.356&\textbf{0.192}&0.461\\
Spice&3.087&\textbf{2.436}&4.776&COMP6-EM0&2.924&\textbf{2.59}&4.496\\
ANI1&0.468&\textbf{0.391}&1.331&COMP6-EM1&2.844&\textbf{2.476}&4.242\\
Qmugs (DFT)&0.288&\textbf{0.192}&0.641&COMP6-EM2&2.899&\textbf{2.498}&4.289\\
Qmugs (PM6)&\textbf{0.277}&0.318&0.532&COMP6-EM3&3.207&\textbf{2.792}&4.827\\
Qmugs (GFN2)&\textbf{0.174}&0.181&0.328&COMP6-EM4&2.707&\textbf{2.412}&4.101\\
QM7X (SE)&0.016&\textbf{0.009}&0.048&COMP6-EM5&2.726&\textbf{2.378}&4.114\\
QM7X (DFT) &0.342&\textbf{0.311}&0.991&COMP6-EM6&2.808&\textbf{2.401}&4.056\\
RevMD17&0.059&0.091&0.257&GDML&\textbf{0.302}&0.581&1.121\\
Transition1X&1.356&\textbf{1.118}&1.881&SN2RXN&\textbf{1.222}&4.382&1.445\\
TMQM&\textbf{12.186}&14.262&18.262&Solvated Peptides&\textbf{1.117}&1.793&2.541\\
Orbnet Denali (DFT)&\textbf{2.113}&2.489&4.471&&&&\\
\midrule
DES370K&1.5173&\textbf{0.9814}&1.1244 & DES5M&0.7809&\textbf{0.3026}&0.5400\\
DESS66*&2.3074&\textbf{0.2307}&0.5650 & DESS66x8*&2.5196&\textbf{0.2256}&0.5831\\
Metcalf*&5.8185&4.7641&\textbf{1.8393}& Splinter&\textbf{1.6576}&1.7597&2.1037\\
\bottomrule
\end{tabular}}
\end{table}

\subsection{Results}
\label{sec:results}
Table \ref{tab:results} reports the mean absolute error (MAE) on the test set for several datasets and our three models. During our analysis, we realized that a bug affected the dataset splits: some were random rather than molecule-based as intended and we are re-training to fix the situation and run multiple times to add error bars subsequently . Accordingly, we will only compare different architectures and discuss differences based on chemical space only after correcting the splits.

Overall on potential datasets, we found that the SchNet architecture is consistently and significantely outperformed by TorchMDNet and DimeNet, in line with previous works since it fails to distinguish between certain molecular geometries.
On the other hand, the performance lead between TorchMDNet and DimeNet varies based on the dataset. Interestingly,  on the interaction datasets, SchNet is more competitive suggesting that potential energy and interaction energy datasets might potentially require different architectural biases. 
Additionally, on a few datasets, all methods fail to reach the chemical accuracy threshold (1 kcal/mol), also suggesting more room for improvements. 


\section{Discussion}
\label{sec:conclusion}
In this work, we presented a curated collection of QM  datasets for training MLIPs. It provides an unprecedented number
of molecular conformations, along with associated energies and atomic forces, and covers various parts of the chemical space. Thereby, it significantly enhances the resources available for research in the field of MLIP towards training universal potentials with greater generalizability and robustness, and will be pushing the frontier on  architectural research.  To facilitate the accessibility to these datasets, we
presented the OpenQDC library designed to efficiently store, load and train with them. The library takes advantage
of the unique features and properties presented by these datasets relative to general 2D or 3D graph libraries that make working with these datasets more challenging.
Furthermore, we presented a number of baseline results on several of the proposed datasets and showed that well-known methods struggle to reach chemical accuracy, demonstrating that more work is needed in the field. \\
\textbf{Limitations}:
Currently, OpenQDC only includes energy and forces labels from the original datasets, neglecting other QM properties that could enhance the quality of MLIP internal representations and their generalization. To solve this, we plan long-term to add QM datasets with densities and Hamiltonians. Additionally, OpenQDC only proposes datasets to train MLIPs but is lacking challenging benchmarks that could focus the efforts of the field. Therefore, we aim to add a few benchmarks subsequently.\\
\textbf{Future work}: We plan to address the limitations above in our future work in the near future. Additionally, we believe OpenQDC opens a future where QM foundation models will be trained with sufficient data to reach the same accuracy as CCSD-like methods and be universally applicable to all organic molecules. We plan on leveraging these collections to bring this future into reality.

\bibliographystyle{unsrtnat}
\bibliography{references} 

\clearpage
\appendix
\section{Appendix}
\subsection{Hyperparameters}
\label{appendix:hyperparameters}
We employ the same hyperparameters for all models (summarized in table \ref{table:TorchMDNet_hyperparameters}) and for all datasets throughout this work. All training runs were performed on 2 Nvidia A100 GPUs with 4 CPU workers per GPU.
The duration of each training run varied on the volume of training data but even in instances with the most extensive data, all training runs were completed in under 52 hours.

\begin{table}[h]
\centering
\caption{Hyperparameters used throughout this work for the various networks architecture. Any hyperparameter not reported is set to his default values.}
\adjustbox{width=\textwidth}{\begin{tabular}{llll}
\toprule

Hyperparameter name & TorchMD-Net& DimeNet& SchNet \\ \midrule
Initial learning rate (LR) & $3 \cdot 10^{-4}$; $10^{-4}$ (S.V2)
& $ 10^{-4}$; $10^{-7}$ (S.V2)
& $ 10^{-4}$; $3 \cdot 10^{-4}$ (S.V2) \\
Batch size & 28& 32& 32 \\
Optimizer & Adam \cite{Kingma2015} & Adam & Adam \\
\midrule
LR Scheduler & ReduceLROnPlateau & ReduceLROnPlateau& ReduceLROnPlateau\\
Decay factor & 0.9 & 0.9 & 0.9 \\
Patience & 10 epochs & 10 epochs& 10 epochs\\
Minimum LR & $10^{-7}$ & $10^{-7}$ & $10^{-7}$\\
\midrule
StochasticWeightAveraging & LR: $10^{-6}$ & LR: $10^{-6}$& LR: $10^{-6}$\\
\midrule
\# Parameters & 1.8 milion & 1.8 milion & 1.2 milion \\
Cut-off & 5 A ; 10 A (S.V2) &5 A ; 10 A (S.V2) &5 A ; 10 A (S.V2)\\
Prediction Heads Hidden Neurons &64 & 128 & 128 \\
Hidden Channels &128 & 128 & 256\\
\# Layers &8 & 4& 8\\
\# Attention Heads &8 & -& -\\
Activation Function & Swish \cite{ramachandran2017searching} & Swish & SoftPlus \cite{schutt2018schnet} \\
\# RBF & 64, Trainable & 6, Trainable & 50\\
Neighbor Embedding & True & - & - \\
\# SBF & -  & 7, Trainable & -\\
\# Filters & -  & - & 128\\
\bottomrule
\end{tabular}}
\label{table:TorchMDNet_hyperparameters}
\end{table}

\subsection{Inference time of the baseline methods}
\label{appendix:inference_time}
\begin{figure}[h]
    \centering 
    \includegraphics[width=\textwidth]{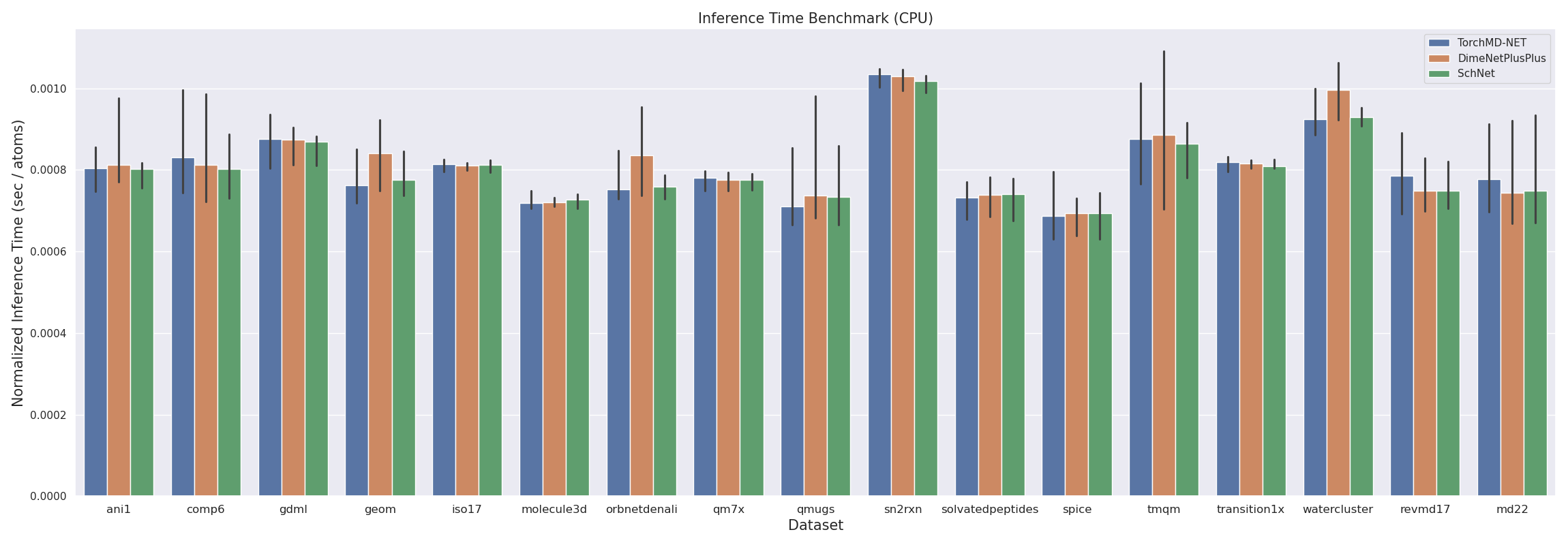}
    \caption{Inference time over the potential datasets.}
    \label{fig:iid_qmugs_qm7x}
\end{figure}



\end{document}